\documentclass[11pt]{article}
\usepackage{graphicx}

\newcommand{\BABARPubYear}    {01}

\newcommand{\BABARProcNumber} {69}
\newcommand{\SLACPubNumber} {9032}

\input pubboard/babarsym
\newcommand\prd[3]   {{{Phys.\ Rev.\ }{\bf D #1} (#2) #3}}
\renewcommand\prep[3]  {{{Phys.\ Rept.\ }{\bf #1} (#2) #3}}
\newcommand\prl[3]   {{{Phys.\ Rev.\ Lett.\ }{\bf #1} (#2) #3}}
\newcommand\ptp[3]   {{{Prog.\ Theor.\ Phys.\ }{\bf #1} (#2) #3}}
\renewcommand{\hepex}[1]{{hep-ex/#1}}
\newcommand\slacrep[2]    {{{\bf SLAC-R-#1} (#2)}}

\begin{document}
{\thispagestyle{empty}

\begin{flushright}
SLAC-PUB-\SLACPubNumber \\
\babar-PROC-\BABARPubYear/\BABARProcNumber \\
%\babar-PUB-\BABARPubYear/\BABARPubNumber \\
%hep-ex/\LANLNumber \\
October, 2001 \\
\end{flushright}

\par\vskip 4cm

% Title of the paper
\begin{center}
  \Large \bf Search for direct $\cal CP$ violation in $\rm
  B\rightarrow K\pi, \pi\pi, K K$, Quasi-Two-Body $\rm B$ decays and
  $\rm B\rightarrow K^*\gamma$ with the \Lbabar~detector at the
  PEP-II collider
\end{center}
\bigskip

\begin{center}
  \large
  S. Menke\\
  Stanford Linear Accelerator Center,
  2575 Sand Hill Road, Menlo Park, CA 94205, USA\\
  (for the \lbabar\ Collaboration)
\end{center}
\bigskip \bigskip

% Abstract
\begin{center}
\large \bf Abstract
\end{center}
A sample of 23 million $\rm B\overline{B}$ events collected with the
\babar\ detector at the PEP-II collider is used in a search for
direct $\cal CP$ violation in charmless two-body $\rm B$ decays, quasi
two-body $\rm B$ decays, and the radiative penguin decays $\rm B
\rightarrow K^*\gamma$. No evidence for direct $\cal CP$ violation is
found in the considered modes and $90\,\%$ confidence level limits are
reported. We also present a limit on the branching fraction of the
decay $\rm B^0\rightarrow \gamma\gamma$.

\vfill
\begin{center}
Contributed to the Proceedings of the International Europhysics Conference 
On High-Energy Physics (HEP 2001),\\ 
7/12/2001---7/18/2001, Budapest, Hungary
\end{center}

\vspace{1.0cm}
\begin{center}
{\em Stanford Linear Accelerator Center, Stanford University, 
Stanford, CA 94309} \\ \vspace{0.1cm}\hrule\vspace{0.1cm}
Work supported in part by Department of Energy contract DE-AC03-76SF00515.
\end{center}

\section{Introduction}
Rare $\rm B$ meson decays are interesting in searches for direct $\cal
CP$ violation because they have significant penguin amplitudes. In the
Standard Model substantial $\cal CP$ violation in $\rm B$ decays could
arise from the interference of penguin and tree
amplitudes~\cite{art:Bander79} and would manifest itself in an
asymmetry of $\rm B$ decay rates:
\begin{equation}\label{eq:Acp}
  {\cal A_{CP}} = \frac{\Gamma(\overline{\rm B}\rightarrow\overline{f})
    - \Gamma({\rm B}\rightarrow{f})}
  {\Gamma(\overline{\rm B}\rightarrow\overline{f})
    + \Gamma({\rm B}\rightarrow{f})}.
\end{equation}
In general the weak phase difference between the $\rm b\rightarrow u$
tree amplitude and the $\rm b\rightarrow s$ or $\rm b\rightarrow d$
penguin amplitude is $\gamma$ or $\gamma+\beta = \pi-\alpha$.
Therefore $\cal A_{CP}$ can be used to constrain the CKM angles
$\alpha$ and $\gamma$ in the phase convention given
in~\cite{art:Wolfenstein83,art:Kobayashi73}.  For pure penguin decays
like $\rm B\rightarrow K^{*}\gamma$ $\cal A_{CP}$ is negligible in the
Standard Model. Extensions of the Standard Model could introduce new
virtual high-mass fermions and bosons in the loop thus providing
additional amplitudes with different phases.  Depending on the model
parameters, $\cal A_{CP}$ may be as large as
$20\,\%$~\cite{art:Kagan98}.

\section{Data Sample}
The data sample used in these analyses was collected with the \babar\ 
detector~\cite{art:BABAR} at the PEP-II $\rm e^+e^-$
collider~\cite{art:PEPII} at SLAC. It corresponds to an integrated
luminosity of $20.7\,{\rm fb}^{-1}$ taken on the $\rm\Upsilon(4S)$
resonance (''on-resonance'') and $2.6\,{\rm fb}^{-1}$ taken at a
center-of-mass energy $40\,{\rm MeV}$ below the $\rm\Upsilon(4S)$
resonance (''off-resonance''), which are used for continuum background
studies.  The on-resonance sample corresponds to $22.6$ million $\rm
B\overline{B}$ pairs. The collider is operated with asymmetric beam
energies, producing a boost ($\beta\gamma = 0.56$) of the
$\rm\Upsilon(4S)$ along the collision axis ($z$). The boost increases
the momentum range of two-body $\rm B$ decay products from a narrow
distribution centered near $2.6\,{\rm GeV}$ to a broad distribution
extending from $1.7\,{\rm GeV}$ to $4.3\,{\rm GeV}$.
  
The \babar\ detector is a spectrometer of charged and neutral
particles and is described in detail in Ref.~\cite{art:BABAR}.
Charged particle (track) momenta are measured in a tracking system
consisting of a 5-layer, double-sided, silicon vertex detector (SVT)
and a 40-layer drift chamber (DCH) filled with a gas mixture of helium
($80\,\%$) and isobutane ($20\,\%$), both operating within a
$1.5\,{\rm T}$ solenoidal magnet. Photons are detected in an
electro-magnetic calorimeter (EMC) consisting of 6580 CsI(Tl)
crystals.  Charged hadron identification is based on the \v Cerenkov
angle~$\theta_{\rm c}$ measured by a unique, internally reflecting \v
Cerenkov ring imaging detector (DIRC).

\section{Event Selection} 
Hadronic events are selected based on track multiplicity and event
topology.  Backgrounds from non hadronic events are reduced by
requiring the ratio of the Fox-Wolfram moments
$H_2/H_0$~\cite{art:Fox78} to be less than $0.95$ and the
sphericity~\cite{art:Wu84} of the event to be greater than $0.01$
(charmless two-body decays~\cite{art:TwoBodyAnal}) or by requiring
$|\cos \theta_T^*| < 0.8$ (${\rm K}^*\gamma$ decays), where
$\theta_T^*$ denotes the angle between the thrust vector of the event
excluding the $\rm B$ daughter candidates and the high energy photon
candidate in the center-of-mass frame.  Candidate tracks are required
to originate from the interaction point, and to have at least 12 DCH
hits and a minimum transverse momentum of $0.1\,{\rm GeV}$. Looser
criteria are applied to tracks forming $\rm K^0_S$ candidates to allow
for displaced decay vertices. Kaon tracks are distinguished from pion
and proton tracks via a likelihood ratio that includes, for momenta
below $0.7\,{\rm GeV}$ ${\rm d}E/{\rm d}x$ information from the SVT
and DCH, and for higher momenta \v Cerenkov angle and number of
photons as measured by the DIRC.
  
Pairs of tracks with opposite charge from a common vertex are combined
to form $\rm K^0_S$, $\phi$, $\rm K^{*0}$ and $\rho^0$ candidates.
Pairs of charged tracks are further combined with a $\pi^0$ or $\eta$
candidate to select $\omega$ or $\eta'$ candidates.  The required mass
ranges for $\phi$, $\omega$, $\eta'$ and $\eta$ candidates are as
follows (in $\rm GeV$): $0.99<m_{{\rm K}^+{\rm K}^-}<1.05$,
$0.735<m_{\pi^+\pi^-\pi^0}<0.83$, $0.93<m_{\eta\pi^+\pi^-}<0.99$,
$0.9<m_{\rho^0\gamma}<1.0$, $0.49<m_{\gamma\gamma}<0.6$.

% K0s ...
${\rm K^0_S}$ candidates should have a mass within $3.5\sigma$ of
the nominal mass, where $\sigma$ is typically $4.3\,{\rm MeV}$ for
two-body $\rm B$ decays, and a proper lifetime significance of at
least 5 for the two-body-analysis. Similar cuts are applied in the
$\rm K^*\gamma$ and the
quasi-two-body-analyses~\cite{art:CharmlessAnal}.
% two-body        m = +-3.5sigma (@sigma = 4.3MeV) tau/sigma > 5
% quasi-two-body  m = +-12MeV    tau/sigma > 3
% K*gamma         m = +-11MeV    0.2cm displacement from IP in transverse plane
  
The $\rho$ mass is required to be in the interval $[0.5,0.995]\,{\rm
  GeV}$.  The ${\rm K}^*$ reconstruction is completed by requiring the
invariant mass of the ${\rm K}\pi$ pairs to be within $\pm100\,{\rm
  MeV}$ of the nominal ${\rm K}^{*0}/{\rm K}^{*+}$ mass, except for
${\rm K}^+\pi^0$ and ${\rm K^0_S}\pi^+$ pairs for the quasi-two-body
analysis where $\pm150\,{\rm MeV}$ are required.
  
${\rm J /\psi}\to\mu^+\mu^-$ candidates are constructed from two
identified muons each with polar angle in the range $[0.3,2.7]\,{\rm
  rad}$ and with invariant mass $3.06\,{\rm GeV} < m_{\mu^+\mu^-} <
3.14\,{\rm GeV}$~\cite{art:JpsiAnal}. The absolute cosine of the
helicity angle of the ${\rm J /\psi}$ decay is required to be less
than 0.9.  ${\rm J /\psi}\to{\rm e}^+{\rm e}^-$ candidates are
constructed from two identified electrons each with polar angle in the
range $[0.41,2.409]\,{\rm rad}$ and with invariant mass $2.95\,{\rm
  GeV} < m_{\mu^+\mu^-} < 3.14\,{\rm GeV}$. The absolute cosine of the
helicity angle of the ${\rm J /\psi}$ decay is required to be less
than 0.8.
  
The $\rm K^*\gamma$ analysis selects high energy photon candidates in
the EMC in the energy range $1.5\,{\rm GeV} < E_\gamma < 4.5\,{\rm
  GeV}$ in the laboratory frame and $2.3\,{\rm GeV} < E_\gamma^* <
2.85\,{\rm GeV}$ in the center-of-mass frame. The candidate must be
isolated by $25\,{\rm cm}$ from any other photon candidate or track
and have a lateral energy profile consistent with a photon shower.
Photons from $\pi^0(\eta)$ are vetoed by requiring that the invariant
mass of the combination with any other photon of energy greater than
$50(250)\,{\rm MeV}$ not lie within the range $115(508)\,{\rm MeV} <
m_{\gamma\gamma} < 155(588)\,{\rm MeV}$.  Similar requirements are
imposed on the high energetic photons in the $\rm B^0\to\gamma\gamma$
analysis~\cite{art:GamGamAnal}.

%pi0, (eta)
%two-body      :       30                  +/-  3 sigma
%quasi-two-body: E_g > 30(100) MeV; m_gg = +/- 15 MeV
%K*gamma               30         ;        +15 - 20 MeV; E_pi0 > 200 MeV
$\pi^0(\eta)$ candidates are formed from pairs of photons with
energies of at least $30(100)\,{\rm MeV}$. The accepted invariant mass
range for $\pi^0$ candidates is typically $[115,150]\,{\rm MeV}$.

%selection of B candidates I m_ES
%two-body                 : m_ES = sqrt((s/2 + p_in*p_B)^2/E_in^2 - p_B^2)     
%charmless quasi-two-body : m_ES = sqrt((s/2 + p_in*p_B)^2/E_in^2 - p_B^2)     
%quasi-two-body           : m_ES = sqrt(s/2 - p_B*^2)     
%K*gamma                  : m_ES = sqrt(s/2 - p_B*^2)     
%gamma gamma              : m_ES = sqrt(s/2 - p_B*^2)     
%
%selection of B candidates II Delta E
%two-body                 : Delta E = sum E_i(pion-mass)^* - sqrt(s)/2
%charmless quasi-two-body : Delta E = (E_in E_B - p_in*p_B - s/2)/sqrt(s)
%                                   = sum E_i^* - sqrt(s)/2
%quasi-two-body           : Delta E = sum E_i^* - sqrt(s)/2
%K*gamma                  : Delta E = sum E_i^* - sqrt(s)/2
%gamma gamma              : Delta E = sum E_i^* - sqrt(s)/2
In all analyses presented here two kinematic variables are used to
select $\rm B$ candidates~\cite{art:BABAR}: $\Delta E = E_{\rm
  B}^*-\sqrt{s}/2$, and $m_{\rm ES} = \sqrt{s/4 - {\bf p}_{\rm
    B}^{*2}}$, where $E_{\rm B}^*$ is the reconstructed energy of the
$\rm B$ candidate in the center-of-mass frame, ${\bf p}_{\rm B}^*$ is
its momentum vector, and $\sqrt{s}$ is the total center-of-mass
energy.  In the two-body analysis pion mass is assumed for both tracks
in the definition of $\Delta E$ whereas the correct mass according to
particle identification is used in the other analyses. Therefore the
$\Delta E$ distribution is peaked near zero for modes with no charged
kaons and shifted on average $-45(-91)\,{\rm MeV}$ for modes with one
(two) charged kaon(s) in the two-body analysis. In the other analyses
$\Delta E$ peaks near zero for all signal modes for true $\rm B$
candidates.  The two-body and charmless quasi-two-body analyses use
Fisher discriminants~$\cal F$~\cite{art:Asner96} built from a nine bin
representation of the energy-flow about the $\rm B$ decay axis and in
case of the charmless quasi-two-body analysis the $\rm\Upsilon(4S)$
and the $\rm B$ helicity angles.  Detailed Monte Carlo (MC)
simulation, off-resonance data, and events in on-resonance $m_{\rm
  ES}$ and $\Delta E$ sideband regions are used to study backgrounds.
The largest source of background is from random combinations of tracks
and neutrals produced in the ${\rm e}^+{\rm e}^-\to{\rm
  q}\overline{\rm q}$ continuum (where $\rm q = u,d,s$, or $\rm c$).
  
\section{Signal Extraction} 
% two-body: unbinned extended maximum Likelihood m_ES, Delta E, Fisher, Theta_C
% J/Psi K:  unbinned extended maximum Likelihood m_ES, Delta E, p_Hadron
% gamma gamma : cut-and-count m_ES    = m_B +/- 2 sigma (sigma = 3.9 MeV) 
%                             Delta_E = 0   +/- 2 sigma (sigma = 73 MeV) 
% K* gamma :unbinned (extended?) maximum Likelihood m_ES
% charmless quasi-two-body: unbinned extended maximum Likelihood m_ES, Delta E, 
%                           m_Hadron(eta',omega,phi,K*,eta), Fisher, 
%                           helicity(phi,omega), Theta_C
Two different analysis strategies are used to extract signals.  A
simpler cut-and-count approach where backgrounds are estimated from
sideband regions in $\Delta E$ and $m_{\rm ES}$ and subtracted from
the signal region gives the number of signal events by counting the
remaining events in the signal region (${\rm B}^0\to\gamma\gamma$).
Alternatively unbinned extended maximum Likelihood fits are performed
and signal as well as background yields and the asymmetry parameters
are determined by the fits. The kinematic variables used in the
likelihoods are: $m_{\rm ES}$ (all analyses); $\Delta E$ (all analyses
but ${\rm K^*}\gamma$); $\cal F$ and $\theta_{\rm c}$ (two-body and
charmless quasi-two-body); $p_{\rm K}$ (${\rm J}/\psi{\rm K}^+$);
$m_{\eta',\omega,\phi,{\rm K^*},\eta}$ and helicity angles for
$\omega$ and $\phi$ (charmless quasi two-body).  The shape and the
fixed parameters for the probability density functions (PDF) are
extracted from signal and background distributions from MC simulation,
on-resonance $\Delta E$-$m_{\rm ES}$ sidebands, and off-resonance
data. The MC resolutions are adjusted by comparisons of data and
simulation in abundant calibration channels with similar kinematics
and topology, such as ${\rm B}\to{\rm D}\pi,{\rm D}\rho$ with ${\rm
  D}\to{\rm K}\pi, {\rm K}\pi\pi$. The \v Cerenkov angle residual
parameterizations are determined from samples of ${\rm D}^0\to{\rm
  K}^-\pi^+$ originating from $\rm D^*$ decays.

\section{Results} 
In table~\ref{tab:results} the results for branching ratios and charge
asymmetries are summarized\footnote{For the $\rm K^0\overline{K}^0$
  mode we assume the Standard Model prediction that $\rm B^0\to K^0_S
  K^0_S$ proceeds through the $\rm K^0\overline{K}^0$ intermediate
  state and use ${\cal B}(\rm K^0\overline{K}^0\to K^0_S K^0_S) =
  0.5$}. Given errors denote statistical and systematic uncertainties,
respectively.  

\begin{table}
  \begin{tabular}{c|cccc}
    Mode & \ ${\cal B}\ (10^{-6})$ & $\cal A_{CP}$ & $\cal A_{CP}$ ($90\,\%$ C.L.) & note \\
    \hline
    $\pi^+\pi^-$            & $4.1\pm1.0\pm0.7$  \\
    ${\rm K}^+\pi^-$        & $16.7\pm1.6\pm1.3$   & $-0.19 \pm 0.10 \pm 0.03$              & $[-0.35,-0.03]$\\
    ${\rm K}^+{\rm K}^-$    & $<2.5$ ($90\,\%$ C.L.) \\
    $\pi^+\pi^0$            & $<9.6$ ($90\,\%$ C.L.) \\
    ${\rm K}^+\pi^0$        & $10.8{{+2.1}\atop{-1.9}}\pm1.0$ & $\phantom{+}0.00 \pm 0.18 \pm 0.04$    & $[-0.30,+0.30]$\\
    ${\rm K}^0\pi^+$        & $18.2{{+3.3}\atop{-3.0}}\pm2.0$ & $-0.21 \pm 0.18 \pm 0.03$              & $[-0.51,+0.09]$\\
    $\overline{\rm K}^0{\rm K}^+$    & $<2.4$ ($90\,\%$ C.L.) \\
    ${\rm K}^0\pi^0$        & $8.2{{+3.1}\atop{-2.7}}\pm1.2$  \\
    ${\rm K}^0\overline{\rm K}^0$  & $<7.3$ ($90\,\%$ C.L.) & & & preliminary \\
    \hline
    ${\rm J}/\psi{\rm K}^+$ & & $\phantom{+}0.004 \pm 0.029 \pm 0.004$ & $[-0.044,+0.052]$ & preliminary \\
    \hline
    $\eta'{\rm K}^+$        & & $-0.11 \pm 0.11 \pm 0.02$              & $[-0.28,+0.07]$\\
    $\omega\pi^+$           & & $-0.01 {{+0.29}\atop{-0.31}}\pm 0.03$    & $[-0.50,+0.46]$\\
    $\phi{\rm K}^+$         & & $-0.05 \pm 0.20 \pm 0.03$              & $[-0.37,+0.28]$\\
    $\phi{\rm K}^{*+}$      & & $-0.43 {{+0.36}\atop{-0.30}} \pm 0.06$   & $[-0.88,+0.18]$\\
    $\phi{\rm K}^{*0}$      & & $\phantom{+}0.00 \pm 0.27 \pm 0.03$    & $[-0.43,+0.43]$\\
    \hline
    ${\rm K}^{*0}_{\rm K^+\pi^-}\gamma$  &$43.9\pm4.1\pm2.7$ & & & preliminary\\
    ${\rm K}^{*0}_{\rm K^0_S\pi^0}\gamma$&$41.0\pm17.1\pm4.2$ & & & preliminary\\
    ${\rm K}^{*+}_{\rm K^0_S\pi^+}\gamma$&$31.2\pm7.6\pm2.1$ & & & preliminary\\
    ${\rm K}^{*+}_{\rm K^+\pi^0}\gamma$  &$55.2\pm10.7\pm4.2$ & & & preliminary\\
%    ${\rm K}^{*+}\gamma$     & $37.7\pm6.3\pm2.0$ & & & preliminary\\
%    ${\rm K}^{*0}\gamma$     & $42.3\pm4.0\pm2.3$ & & & preliminary\\
    ${\rm K}^{*}\gamma$      & & $-0.035 \pm 0.076 \pm 0.012$           & $[-0.16,+0.09]$ & preliminary \\
    \hline $\gamma \gamma$ & $<1.7$ ($90\,\%$ C.L.)
  \end{tabular}%
  \caption{Branching ratios and $\cal CP$ violating charge asymmetries in 
    rare two-body and quasi-two-body $\rm B$
    decays.\label{tab:results}}
\end{table}
$90\,\%$ C.L. upper limits for the branching ratios are given in cases 
where no signal is observed or the significance of the signal is less than
$4\sigma$.       

Systematic uncertainties arise from: imperfect knowledge of the PDF
shapes, uncertainties in the detection efficiencies, and potential
charge bias in track reconstruction and particle identification. For
most of the measurements, the PDF shapes contribute the largest
systematic error. These uncertainties are often dominated by the
statistical error on the used control sample. Due to the high
statistics in the $\rm J/\psi K^+$ mode greater care is taken of
possible systematic effects.  The fake asymmetry due to the different
probability of interaction of $\rm K^+$ and $\rm K^-$ in the detector
material before the DCH is estimated to be $-0.0039$. We correct $\cal
A_{CP}$ in this mode by this number and add $100\,\%$ of its magnitude
to the systematic error.

\section{Conclusions} 
We have measured branching fractions for the rare charmless decays
$\rm B^0\to\pi^+\pi^-$, $\rm B^0\to K^+\pi^-$, $\rm B^+\to K^+\pi^0$,
$\rm B^+\to K^0\pi^+$, and $\rm B^0\to K^0\pi^0$, and set upper limits
on $\rm B^0\to K^+K^-$, $\rm B^+\to\pi^+\pi^0$, $\rm
B^+\to\overline{K}^0K^+$, and a preliminary upper limit on $\rm B^0\to
K^0\overline{K}^0$. We also report preliminary branching ratios for
the radiative penguin decays $\rm B^{0(+)}\to K^{*0(+)}\gamma$ and set
an upper limit on $\rm B^0\to\gamma\gamma$.

We found no evidence for direct $\cal CP$ violation in the considered
modes and $90\,\%$ confidence level limits are reported.

\end{document}